\documentclass{article}

\usepackage{PRIMEarxiv}
\usepackage[normalem]{ulem}
\usepackage[utf8]{inputenc} 
\usepackage[T1]{fontenc}    
\usepackage{hyperref}       
\usepackage{url}            
\usepackage{booktabs}       
\usepackage{amsfonts}       
\usepackage{nicefrac}       
\usepackage{microtype}      
\usepackage{lipsum}
\usepackage{fancyhdr}       
\usepackage{graphicx}       
\graphicspath{{media/}}     

\title{FE-TCM: Filter-Enhanced Transformer Click Model for Web Search
}

\author{
  Yingfei Wang \\
  College of Computer Science and Engineering \\
  North Minzu University\\
  Yinchuan\\
  \texttt{20217426@stu.nmu.edu.cn} \\
   \And
  Jianping Liu \\
  College of Computer Science and Engineering \\
  North Minzu University\\
  Yinchuan\\
  \texttt{liujianping01@nmu.edu.cn} \\
  \And
  Jian Wang\\
  Agricultural Information Institute \\
  Chinese Academy of Agricultural Sciences\\
  Beijing\\
  \texttt{wangjian01@caas.cn} \\
  \And
  Xiaofeng Wang\\
  College of Computer Science and Engineering \\
  North Minzu University\\
  Yinchuan\\
  \texttt{xfwang@nmu.edu.cn} \\
  \And
  Meng Wang \\
  College of Computer Science and Engineering \\
  North Minzu University\\
  Yinchuan\\
  \texttt{20217441@stu.nmu.edu.cn} \\
  \And
  Xintao Chu \\
  College of Computer Science and Engineering \\
  North Minzu University\\
  Yinchuan\\
  \texttt{xtchu@stu.nun.edu.cn} \\
}

\begin{document}
\maketitle

\begin{abstract}
Constructing click models and extracting implicit relevance feedback information from the interaction between users and search engines are very important to improve the ranking of search results. Using neural network to model users' click behaviors has become one of the effective methods to construct click models. In this paper, We use Transformer as the backbone network of feature extraction, add filter layer innovatively, and propose a new Filter-Enhanced Transformer Click Model (FE-TCM) for web search. Firstly, in order to reduce the influence of noise on user behavior data, we use the learnable filters to filter log noise. Secondly, following the examination hypothesis, we model the attraction estimator and examination predictor respectively to output the attractiveness scores and examination probabilities. A novel transformer model is used to learn the deeper representation among different features. Finally, we apply the combination functions to integrate attractiveness scores and examination probabilities into the click prediction. From our experiments on two real-world session datasets, it is proved that FE-TCM outperforms the existing click models for the click prediction.
\end{abstract}

\keywords{click model \and click prediction \and web search \and transformer}

\section{Introduction}
Search result ranking is one of the major concerns in search engine researches. Click models construction which aim to improving ranking performance with the help of implicit relevance feedback information from the interaction between users and search engines has been paid much attention.

Traditional click models are based on the Probabilistic Graph Model (PGM) framework\cite{koller2009probabilistic}, where user behaviors are represented as a sequence of observable variables (e.g., clicks) and hidden variables (e.g.,examination, skip, relevance,etc.). PGM-based click models have strong explanatory. However, PGM-based click models need to manually design the dependencies among binary variables, which may over-simplify user behaviors\cite{borisov2016neural}. With the development of deep learning, researchers construct click models by Neural Network (NN). NN-based click models can improve the accuracy of user behavior prediction by enhancing expression abilities and allowing flexible dependencies. Numerous studies show that the NN-based click models outperform the PGM-based click models for the click prediction\cite{borisov2016neural,chen2020context,borisov2018click,lin2021graph}.

Transformer\cite{vaswani2017attention} is the latest state of the art in sequence-to-sequence learning, and in many natural language processing tasks, such as text classification and context prediction, Transformer performs well\cite{li2021act,wang2021transformer}. Some recent works have started to use Transformer to construct click models. Li et. al.\cite{li2020interpretable} proposed the InterHAt model, which uses efficient attention aggregation strategy to learn high-order feature interaction and realize click prediction. Bisht et. al.\cite{bisht2022v}proposed v-TCM model, which is based on Transformer and also learns user behaviors from vertical information.

However, user click behaviors are usually complicated and noisy, and logged user behavior data are in essence noisy\cite{agichtein2006improving,said2012users}. Previous studies have shown that deep neural networks tend to overfit on noisy data\cite{caruana2000overfitting,lever2016points}. When the logged user behavior data contains noise, the performance of Transformer based on self-attention mechanism will be degraded because it pays attention to all feature embedding of sequence modeling\cite{zhou2022filter}.

Considering the above issues, we introduce filtering algorithms in the field of digital signal processing into click models, and use filtering algorithms to attenuate the noise, and reduce the influence of noise for sequence data\cite{rabiner1975theory}. We suspect that when the sequence data was denoised, it will be easier to capture sequence user behaviors.

In this paper, we propose a novel Filter-Enhanced Transformer Click Model (FE-TCM) for web search. Firstly, following the examination hypothesis\cite{chuklin2015click}, we model attractiveness estimator and examination predictor respectively, and output attractiveness scores and examination probabilities. The Transformer with multi-head self-attention mechanism is used for feature learning, and extract the dependencies between different features in user behavior sequence. Secondly, we add learnable filters between embedding layer and backbone network layer to reduce the influence of noise for sequence data. Finally, we combine attraction scores and examination probabilities through the combination layer to complete user click prediction task. 

The main contributions of this paper are as follows: Firstly,we introduce filtering algorithms into the click models to reduce the influence of noise on user behavior data. Secondly, we propose a Filter-Enhanced Transformer Click Model (FE-TCM). FE-TCM uses the Transformer with multi-head self-attention mechanism, and this model is used for learning the position bias from user logs. Finally, From our experiments on two real-world session datasets Yandex and TREC2014, the proposed FE-TCM achieves significantly better performance than existing click models in click prediction task.

\section{Related work}
\subsection{Click model}
In recent years, researchers have proposed numerous click models to describe users' search behaviors in search engines. Traditional click models are based on PGM framework. PGM-based click models treat user's search behaviors as a sequence of observable and hidden events, and manually design the dependencies between these binary events, which is flexible and explanatory. Craswell et. al.\cite{craswell2008experimental} first propose the cascade model (CM), which assumes that users scan each search result from top to bottom until the first click. In the CM, users always leave the search engine result page (SERP) after first click and never return. However, User Browsing Model (UBM)\cite{dupret2008user}, Dynamic Bayesian Network (DBN)\cite{chapelle2009dynamic}, Dependent Click Model (DCM)\cite{guo2009efficient} and Click Chain Model (CCM)\cite{guo2009click} have been proposed to overcome the limitation of CM.

Due to the limited expression ability of PGM, PGM-based click models only model the important factors that affect search behaviors. If a more complex search scenario is to be considered in the PGM, the model will introduce more binary variables, and also need to manually design the dependencies among these new variables, which will cause many difficulties in the iteration and calculation of click models\cite{liu2017user}. Therefore, Researchers try to construct click models by neural network. Borisov et. al.\cite{borisov2016neural} first propose the neural click model (NCM). User behaviors are not binary events in traditional methods, but are modeled as a sequence of vector states, which are iteratively updated with the interaction between users and search engines. Click Sequence Model (CSM)\cite{borisov2018click} follows the encoder-decoder architecture, mainly focusing on the prediction of user click sequences in search engines. Chen et. al.\cite{chen2020context} propose a context-aware model (CACM), It models the session context with an end-to-end neural network and jointly learns the relevance scores and the examination probability of a specific document respectively. Lin et. al.\cite{lin2021graph} propose a graph-enhanced model (GraphCM), which combines intra-session and inter-session information by applying graph neural network and neighbor interaction technology, and effectively alleviates the data sparsity and cold start problem.

Our proposed model applies the latest Transformer and incorporates a filter layer to filter the log noise. Compared with the existing click models, its performance has reached the most advanced level for the click prediction.

\subsection{Transformer}
Transformer is the latest state of the art in sequence-to-sequence learning. It is based on a self-attention mechanism, which effectively alleviates the time dependence of RNN on long sequences. SASRec\cite{kang2018self} and BERT4Rec\cite{sun2019bert4rec} have proved the effectiveness of Transformer in sequence problems, so some recent works have started adapting transformer for constructing click models. Li et. al.\cite{li2020interpretable} propose the InterHAt model, which uses efficient attention aggregation strategy to learn high-order feature interaction for the task of click prediction. Bisht et. al.\cite{bisht2022v} propose the v-TCM model, which is based on the Transformer structure. v-TCM incorporates the vertical information type of each document in SERP (vertical bias) as an additional input to the encoder of a multi-head self-attention based transformer, apart from the query and the ranked search engine results (position bias). Zhou et. al.\cite{RN210} propose an advertising click rate prediction model SACSN based on improved Transformer, which not only effectively models users' historical interests, but also considers the relationship with target advertisements. Chen et. al.\cite{chen2019behavior} propose BST model, which uses Transformer to capture the sequential signals behind users behavior sequence, and extensive experiments prove the superiority of Transformer in modeling the user behavior sequence.

Due to the superiority of Transformer in processing sequence data, we use Transformer to model users' click sequence, and learn the deeper representation of each feature by capturing the relationship with other features in the behavior sequence.

\section{Preliminaries}
In this section, we formulate the click model problem, and then introduce the Fourier transform.
\subsection{Problem formulation}
Users clicking behaviors occur in each search session, and the search session $ S $ can be regarded as a sequence of queries $ Q_{n}=[q_{1},q_{2},…,q_{n}]$ submitted by users. After users submit a query $q_{i} $ to the search engine, the search engine will return a ranked list of documents $D_{i} =[d_{i,1},d_{i,2},…,d_{i,n}]$. Each document $d_{i,j} $ contains two attributes: the unique URL identifier $u_{i,j} $ and the ranking position $p_{i,j}$. The documents are ranked according to their relevance to the query  and presented to users in the form of ranked list. Users examine any document and decides whether to click, if $c_{i,j}=1$, it is regarded as a click, and 0 if not. We can define the problem of click model as follows:

Given the user's queries $ Q=[q_{1},q_{2},…,q_{n}]$, documents $D=[d_{1,1},d_{1,2},…,d_{n,m}]$ and clicks $C=[c_{1,1},c_{1,2},…,c_{n,m-1}]$, for the m-th document $d_{n,m}$ in the n-th query $q_{n} $ of session $S$. the task of our model is to predict whether users will click the document (i.e. the click variable $c_{n,m}$ ). 

In this paper, we model examination predictor and attractiveness estimator respectively, and output attractiveness scores and examination probabilities. In the attractiveness estimator, we take user queries $q$ , documents $d$, click variables $c$ and ranked position $p$ as inputs. In the examination predictor, considering that the user's examination operation is only affected by his/her operation on the previous results in the current query, we take click variables $c$  and ranked position $p$ as inputs. Finally, we combine the attraction score $\mathcal{A} _{n,m}$ and examination probability $\mathcal{E} _{n,m} $ to output the final click probability.

\subsection{Fourier transform}
Discrete Fourier Transform (DFT) is essential in the field of digital signal processing\cite{soliman1990continuous}. Discrete Fourier Transform (DFT) is a discrete form of continuous Fourier Transform in both time domain and frequency domain. In practical applications, the Fast Fourier Transform (FFT) is usually used to efficiently calculate DFT\cite{van1992computational}. Given a sequence $\left \{  x_{k} \right \} $ with $k\in [0,N-1]$, the sequence is converted into frequency domain by 1D DFT:
\begin{equation}
x_{n}=\sum_{k=0}^{N-1}x_{k}e^{-i\frac{2\pi}{N}kn} ,n=0,1,\dots,N-1
\end{equation}

Given the $DFTx_{n} $, We can convert it into the original sequence by inverse DFT (IDFT):
\begin{equation}
x_{k}=\frac{1}{N}x_{n}e^{\frac{2\pi i}{N}nk }  
\end{equation}

\section{Model framework}\label{sec:filetypes}
As shown in Figure 1, we will introduce the overall framework of FE-TCM.
\begin{figure}
  \centering
  \includegraphics[width=1.0\textwidth]{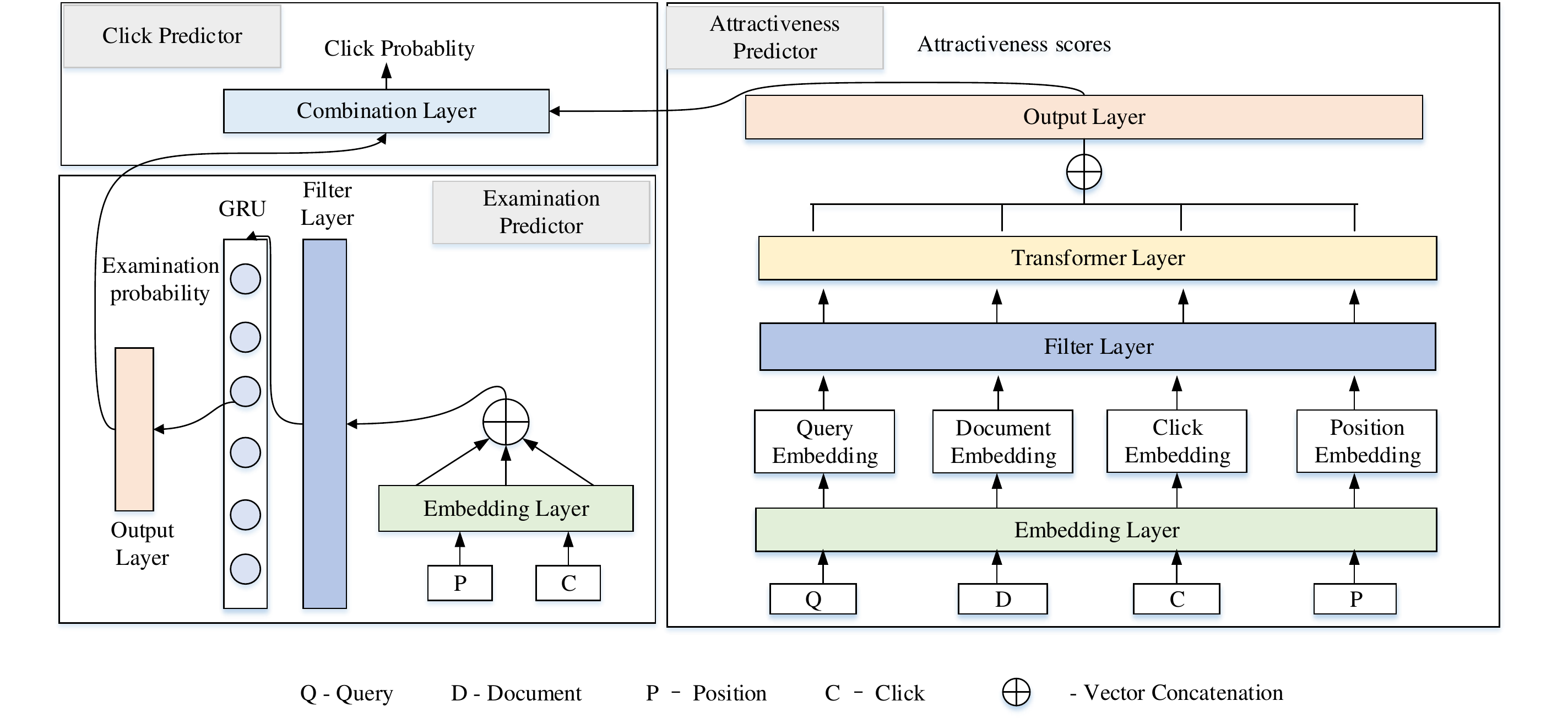}
  \caption{Overall framework of FE-TCM. FE-TCM consists of an attractiveness estimator and a examination predictor. The attractiveness predictor is used to estimate the attraction score $\mathcal{A} _{n,m}$. The examination predictor is used to predict the examination probability $\mathcal{E} _{n,m} $. FE-TCM integrates $\mathcal{E} _{n,m}$  and $\mathcal{A} _{n,m}$ through a combination layer to predict user click behaviors.}
\label{fig:ka}
\end{figure}

\subsection{Embedding layer}
The first component is the embedding layer, which embeds all input features into fixed-size low-dimensional vectors. FE-TCM takes query $q$ , document $d$, user click $c$ and ranked position $p$ as inputs, these original ID features are transformed into a high-dimensional sparse features via one-hot encoding, then we convert high-dimensional sparse vector into low-dimensional dense vector through embedding layer:
\begin{equation}
v_{q}=Emb_{q}(q),v_{d}=Emb_{d}(d),v_{c}=Emb_{c}(c),v_{p}=Emb_{p}(p)
\end{equation}

Where $Emb_{*} \in \mathbb{R}^{N_{*} \times l_{*}} $, $*\in \left \{ q,d,c,p \right \} $. $N_{*}$ and $l_{*} $ denote the input feature size and embedding size. 

\subsection{Filter layer}
Based on the embedding layer, we develop the input features of transformer encoder by stacking multiple learnable filter blocks. In the filtering layer, we first perform filtering operation for each dimension of features in the frequency domain, in order to avoid over-fitting of FE-TCM, we perform skip connection and layer normalization. Given the representation matrix $F^{l}  \in \mathbb{R}^{n \times d}$ , we first perform FFT along its dimension to transform it into the frequency domain:
\begin{equation}
X^{l}=\mathcal{F} (F^{l} )
\end{equation}

Where $\mathcal{F} ( \cdot )$ represents one-dimensional FFT, and $X^{l} $ represents frequency spectrum of $F^{l}$. Then, We can modulate the spectrum by multiplying a learnable filter $W=\mathbb{C} ^{n\times d}$:
\begin{equation}
\widetilde{X}^{l}=W\odot X^{l}    
\end{equation}

Finally, we adopt the inverse FFT to transform the modulated spectrum back to the time domain and update the sequence representations: 
\begin{equation}
\widetilde{F}^{l}\gets F^{-1} (\widetilde{X}^{l}) 
\end{equation}

Where $F^{-1} (\cdot ) $ denotes the inverse 1D FFT, which converts the complex tensor into a real number tensor. Through FFT and inverse FFT operations, the noise of recorded data can be effectively reduced, thus obtaining more pure feature embedding.

In order to alleviate the problem of gradient vanishing and unstable training problems, we also incorporate the skip connection, layer normalization and dropout operations. LayerNorm represents layer normalization, which is mainly used to speed up the convergence of the model; Dropout is a random inactivation, which is used to prevent over-fitting, and the skip connection is adopted to reduce the learning load of the model. The formula is as follows:
\begin{equation}
\widetilde{F}^{l}=LayerNorm(F^{l}+Dropout(\widetilde{F}^{l}))  
\end{equation}

\subsection{Examination predictor}
Following the examination hypothesis\cite{chuklin2015click}, the probability that users click the document depends on the user examining the document and considering it relevant to the query. The examination predictor aims to predict whether the user will continue to examine the document based on his/her session context. In FE-TCM, we follow the hypothesis of previous work\cite{chen2020context,lin2021graph}, the user's examination behavior is only affected by his/her operation on previous documents. Therefore, for the current document $d_{i,j}$, we apply a session-level GRU\cite{chung2014empirical} to encode the ranking position $p$ and historical clicks $c$ in the same session.
\begin{equation}
x_{i,j}=[v_{p{\scriptsize i,j}} \oplus v_{c{\scriptsize i,j}}]
\end{equation}
\begin{equation}
\mathcal{E}_{i,j}^{’}  =GRU(x_{1,1},\dots ,x_{i,j}) 
\end{equation}

Finally, we apply a linear layer and a sigmoid function to realize normalization and output the final examination probability $\mathcal{E}$.

\begin{equation}
\mathcal{E}_{i,j} =Sigmoid(Linear(\mathcal{E}_{i,j}^{’}))
\end{equation}

\subsection{Attractiveness estimator}
Next, we will introduce the attractiveness estimator. The attractiveness estimator aims to estimate the attractiveness of each document $d_{i,j} $ to the user who issues the query $q_{i} $. After the filter layer, FE-TCM embeds all the features as inputs of Transformer encoder. Transformer learns the deeper representation of each feature by capturing the relationship with other features in the behavior sequence.

\textbf{Multi-head self-attention layer.}
The embedding vector $E$ is calculated by multi-head self-attention, and the process is as follows: $E$ is converted into query vector $Q=EW_{i}^{Q}$, keyword vector $K=EW_{i}^{K}$ and value vector $V=EW_{i}^{V}$ by $W_{i}^{Q}$, $W_{i}^{K}$ and $W_{i}^{V}$ respectively, where $W_{i}^{Q}$, $W_{i}^{K}$ and $W_{i}^{V}$ are parameter matrices. Then we compute the dot products of the query with all keys, divide each by $\sqrt{d_{k}}$,and apply a softmax function to obtain the weights on the values. 
\begin{equation}
Attention(Q,K,V)=softmax(\frac{QK^{T}}{\sqrt{d}})V
\end{equation}

Multi-head attention allows the model to jointly attend to information from different representation subspaces at different positions.Following \cite{vaswani2017attention}, we use the multi-head attention:
\begin{equation}
head_{i} =Attention(EW_{i}^{Q},EW_{i}^{K},EW_{i}^{V})
\end{equation}
\begin{equation}
S=concat(head_{1},\dots,head_{h})W^{O} 
\end{equation}

Where $head_{i}$ represents the i-th self-attention. In our task, the self-attention operation takes the filtered embedding of query vector, document vector, ranked position vector and target click vector as inputs, then converts them into three matrices through linear projection, and feeds them to the attention layer.

\textbf{Point-wise Feed-Forward Networks.}
The essence of multi-head attention is linear transformation, we incorporate Point-wise Feed-Forward Networks (FFN) to further enhance the model with non-linearity, which is defined as follows.
\begin{equation}
F=FFN(S)=ReLU(SW^{(1)}+b^{(1)})W^{(2)}+b^{(2)}  
\end{equation}

Where $F$ is the output vector of the FFN, and $W^{(1)}$,$W^{(2)}$,$b^{(1)}$,$b^{(2)}$ are learnable parameters.

Finally, we concatenate the query vector $Q$, the document vector $D$, the click vector $C$ and the position vector $P$ output by Transformer, and generate the final attraction scores $\mathcal{A} $ through a linear layer and a sigmoid function.
\begin{equation}
output=[q\oplus d\oplus c\oplus p]
\end{equation}
\begin{equation}
\mathcal{A}=Sigmoid(Linear(output))
\end{equation}

\subsection{Combination}
The click prediction module combines the examination probability $\mathcal{E}$ and attraction score $\mathcal{A}$ to output the final click probability. We have implemented five different combination functions, as shown in Table 1.

For $mul$ function, we multiply the attraction score $\mathcal{A}$ with the examination probability $\mathcal{E}$ directly. The $exp\_mul$ follows the examination hypothesis and increases the model capacity by adding learnable parameters. Linear function and nonlinear function further discuss the relationship between $\mathcal{E}$ and $\mathcal{A}$. $Sigmoid\_log$ function uses sigmoid function and logarithmic function, and finally obtain a simple formula. Among the five combination functions, only $mul$, $exp\_mul$ and $sigmoid\_log$ support the examination hypothesis.

\subsection{Loss function}
In order to learn the weights and parameters of the model, We take the cross entropy loss function as the objective function, and the formula is expressed as:
\begin{equation}
L=-\frac{1}{N}\sum_{i}^{} \sum_{j}^{}(C_{i,j} logP_{i,j}+(1-(C_{i,j})log(1-P_{i,j})) 
\end{equation}

Where $n$ is the number of training samples. $C_{i,j}$ and $P_{i,j}$ denote the true click signal and the predicted click probability of the r-th result in the i-th query session in the testing set. 

\begin{table*}
  \centering
  \begin{center}
  \caption{Combination functions.}
  \label{tab:freq}
  \begin{tabular}{ccl}
    \toprule
    Function      & Forluma        & Support E.H.? \\
    \midrule
    mul           & $\mathcal{C} =\mathcal{A}\cdot \mathcal{E}$  
    & Yes\\
    exp\_mul       & $\mathcal{C} =\mathcal{A^{\lambda } }\cdot \mathcal{E^{\mu } }  $               & Yes\\
    sigmoid\_log   & $\mathcal{C} =4\sigma (log(\mathcal{A})\cdot \sigma\_log(\mathcal{E})) =4\mathcal{A}\mathcal{E}/(\mathcal{A}+1)(\mathcal{E}+1) $               & Yes\\
    linear        & $\mathcal{C} =\alpha \cdot \mathcal{A} +\beta \cdot \mathcal{E} $               & No\\
    nonlinear     & $\mathcal{C} =MLP(\mathcal{A},\mathcal{E})$               & No\\
  \bottomrule
\end{tabular}
\end{center}
\end{table*}

\section{Experiment}
In this section, we conduct experiments to answer the following questions:

\textbf{RQ1} Compared with baseline click models, Does FE-TCM achieve the best performance in click prediction task on Yandex and TREC2014 two datasets?

\textbf{RQ2} Which combination function performs best when integrating attraction scores and examination probabilities?

\textbf{RQ3} What is the influence of different components in FE-TCM?

\textbf{RQ4} Does the learnable filters improve the model performance?

\subsection{experimental setup}
\subsubsection{Implementation details} 
The operating system used is Windows, the GPU is NVIDIA TITAN V. We train our model with Adam optimizer. We give detailed model parameters in Table 2.

\begin{table}
  \centering
  \begin{center}
  \caption{Configuration of FE-TCM.}
\def~{\hphantom{0}}
  \begin{tabular}{lccc}
  \toprule
    embedding size      & 64    & batch size      & 64\\
    hidden size         & 64    & learning rate   & 0.001\\
    head number         & 8     & dropout         & 0.5\\
    transform block     & 1     & weight decay    & $10^{-5}$\\
  \bottomrule
  \end{tabular}
  \label{tab:kd}
  \end{center}
\end{table}

\subsubsection{Dataset} 
We conducted experiments on two public session datasets. The statistics of the datasets
can be found in Table 3. 

\begin{table}
  \begin{center}
  \caption{The dataset statistics.}
\def~{\hphantom{0}}
  \begin{tabular}{lccc}
  \toprule
    Dataset     & Query     & Session     & Search engine \\
  \midrule
    Yandex      & 376,965   & 200,000     & Yandex \\
    TREC2014    & 5443      & 1257        & Irdia   \\
  \bottomrule
  \end{tabular}
  \label{tab:kd}
  \end{center}
\end{table}

(1)Yandex\footnote{https://www.kaggle.com/c/yandex-personalized-web-search-challenge}: The dataset includes user sessions extracted from Yandex logs, with user ids, queries, query terms, URLs, their domains, URL rankings and clicks. 

(2)TREC2014\footnote{https://trec.nist.gov/data/session2014.html}: The dataset includes queries, URLs, URL ranking position, clicks, and the time spent by users reading the web page corresponding to each click.

Due to the limitation of memory, we randomly sample the sessions in the Yandex dataset. All datasets are divided into training set, validation set and test set according to the ratio of 8:1:1.

\subsubsection{Evalution metric} 
We compare our proposed model with other baseline models in the click prediction, and we report the perplexity (PPL) and log-likelihood (LL) of each model in the paper. Perplexity and log-likelihood are defined as follows:
\begin{equation}
PPL@r=2^{-\frac{1}{N} {\textstyle \sum_{i=1}^{N}C_{i,r}logP_{i,r}+(1-C_{i,r})log(1-P_{i,r})}} 
\end{equation}
\begin{equation}
LL=\frac{1}{MN}\sum_{i=1}^{N}\sum_{j=1}^{M} C_{i,j}logP_{i,j}+(1-C_{i,j})log(1-P_{i,j})
\end{equation}

Where $r$ represents the ranking position on the search engine results page, $N$ represents the number of sessions, $M$ represents the number of results in a query, $C_{i,r}$ is the actual click, and $P_{i,r}$ is the predicted click probability of the r-th document of the i-th query in the test set. We average the perplexity values of all positions to get the overall perplexities of the model. The lower value of perplexity and the higher value of log-likelihood correspond to the better click prediction performance.

\subsubsection{Baselines} 
The existing click models can be divided into two class: PGM-based click models and NN-based click models\cite{bidekani2020ensemble}. For PGM-based click models, we choose five representative models: UBM, DCM, DBN, SDBN and CCM, all of which open-source implementations are available\footnote{https://github.com/markovi/PyClick}. For NN-based click models, we choose NCM and GraphCM as baseline models.

\subsection{Performance comparison (answer RQ1)}
We perform click prediction task for each click model to compare the performance. The results are shown in Table 4, from which we can get the following observations:

\begin{table}
\centering
\caption{Overall performance of each click model. The best results are shown in bold, and the second best results are underlined. The lower value of PPL and the higher value of LL correspond to the better click prediction performance.}
\begin{tabular}{ccccc} 
\hline
Model   & \multicolumn{2}{c}{Yandex}         & \multicolumn{2}{c}{TREC2014}        \\ 
\cline{2-5}
        & LL               & PPL             & LL               & PPL              \\ 
\hline
CCM     & -0.2975          & 1.3393          & -0.3632          & 1.3081           \\
DCM     & -0.3100          & 1.3480          & -0.3728          & 1.2702           \\
DBN     & -0.2980          & 1.3354          & -0.3633          & 1.2847           \\
SDBN    & -0.3034          & 1.3353          & -0.3707          & 1.2858           \\
UBM     & -0.2530          & 1.332           & -0.1732          & 1.2236           \\
NCM     & -0.2333          & 1.2851          & -0.1652          & 1.1880           \\
GraphCM & \uline{-0.2192}  & \uline{1.2652}  & \uline{-0.1529}  & \uline{1.1721}   \\
FE-TCM  & \textbf{-0.2106} & \textbf{1.2543} & \textbf{-0.1475} & \textbf{1.1689}  \\
\hline
\end{tabular}
\end{table}

(1)Among all PGM-based click models, UBM has the advantages of simple structure, low computational cost, and always achieves good experimental results, and performs best among all PGM-based click models.

(2)NCM, GraphCM and FE-TCM are superior to all PGM-based click models in click prediction task. GraphCM performs best among all baseline models. GraphCM combines intra-session and inter-session information by using graph neural network (GAT) and neighbor interaction techniques, so GraphCM can better capture more subtle patterns in user click behaviors.

(3)Our proposed FE-TCM obviously outperforms all baseline models. This improvement proves the superiority of Transformer with multi-head self-attention mechanism in processing sequence data, and the effectiveness of the learnable filters in alleviating the impact of noise in click models.

\subsection{Combination function (answer RQ2)}
 We study the influence of different combination functions on the performance of FE-TCM. From Table 5, we can obtain that $exp\_mul$ function can achieve the best results in the combination layer. Compared with $mul$, $exp\_mul$ and $sigmoid\_log$ functions, $linear$ and $nonlinear$ functions have relatively poor results because they do not support the examination hypothesis.

 \begin{table}
\centering
\caption{Performance comparison of models with different combination functions}
\begin{tabular}{ccccc} 
\hline
Combination function & \multicolumn{2}{c}{Yandex}         & \multicolumn{2}{c}{TREC2014}        \\ 
\cline{2-5}
                     & LL               & PPL             & LL               & PPL              \\ 
\hline
FE-TCM $linear$        & -0.2328           & 1.2845          & -0.1672          & 1.1980            \\
FE-TCM $nonlinear$     & -0.2410           & 1.2917          & -0.1754          & 1.2046           \\
FE-TCM$ sigmoid\_log$  & -0.2199          & 1.2675          & -0.1494          & 1.1712           \\
FE-TCM $mul $          & -0.2179          & 1.2654          & -0.1487          & 1.1706           \\
FE-TCM $exp\_mul  $    & \textbf{-0.2106} & \textbf{1.2543} & \textbf{-0.1475} & \textbf{1.1689}  \\ 
\hline

\end{tabular}
\end{table}

\subsection{Ablation study (answer RQ3, RQ4)}
In order to study the contribution of each module to the overall performance of FE-TCM, and to verify whether the learnable filters can improve the performance of the model, we conducted several comparative experiments. The results are presented in Table 6. 

\begin{table}
\centering
\caption{Analysis of key components of FE-TCM}
\begin{tabular}{ccccc} 
\hline
Method                                     & \multicolumn{2}{c}{Yandex} & \multicolumn{2}{c}{TREC2014}  \\ 
\cline{2-5}
                                           & LL      & PPL              & LL      & PPL                 \\ 
\hline
Transformer+LSTM                           & -0.2238 & 1.2736           & -0.1498 & 1.1718              \\
Transformer+GRU                            & -0.2229 & 1.2725           & -0.1488 & 1.1706              \\
Transformer+Attra+Exam\_Filter+GRU         & -0.2218 & 1.2658           & -0.1496 & 1.1714              \\
Transformer+Attra\_Filter+Exam+GRU         & -0.2228 & 1.2721           & -0.1477 & 1.1691              \\
Transformer+Attra\_Filter+Exam\_Filter+GRU & -0.2106 & 1.2543           & -0.1475 & 1.1689              \\ 
\hline

\end{tabular}
\end{table}

 It can be seen from Table 6, compared with long short-term memory (LSTM), the LL value and PPL value can be increased by about 0.1\% by using gated recurrent unit (GRU) in the examination predictor; We also try to use Transform in the examination predictor, but the experiment result is a little worse than using GRU. We assume it may be that there are fewer input features in this module, so using GRU with simple structure can achieve better results. Besides, by adding filter layers to the examination predictor and attraction estimator, LL and PPL are both improved, which indicates that the learnable filters can improve the performance of GRU and Transformer structures. When the filter layer are added to both examination predictor and attraction estimator , the overall PPL value of the model is increased by 1.8\% on Yandex dataset, and the PPL value is 1.2543, LL value is increased by 1.3\%, and LL value is -0.2106; For the TREC2014 dataset, the overall PPL value of the model increased by 0.17\% and the PPL value is 1.1689, the LL value increased by 0.13\%, and the LL value is -0.1475. The results prove the effectiveness of adding learnable filters between the embedded layer and backbone network layers.

\section{Discussion and future work}
\subsection{Discussion}
In this paper, we propose a new Filter-Enhanced Transformer Click Model (FE-TCM) for web search. FE-TCM consists of an attraction estimator and a examination predictor for click prediction. Specifically the attraction estimator uses Transformer to extract the dependence of different features in users behavior sequence. In order to achieve the best performance, we further study five combination functions to integrate attraction and examination into click prediction.

Extensive experiments are conducted on two real-world session datasets, By answering the four research questions, we find that: 1) Compared with the existing advanced click models, our proposed model outperforms in click prediction task due to the application of transformer and learnable filters. 2) Transformer with multi-head self-attention mechanism is more suitable for capturing sequence signals, and can better learn the position bias from user logs. This structure can be used as the most effective feature extraction network for click prediction. 3) Click models are mainly based on logged user behavior data to fit model parameters, which usually contains noisy interactions. Ablation study shows that filtering algorithms from digital signal processing can effectively alleviate the influence of noise in click models, and proves that adding learnable filters between the embedding layer and the backbone network layer can significantly improve the performance of the model. 4) Among the five combination functions, $exp\_mul$ function has the best comprehensive performance because it supports the examination hypothesis and has more learnable parameters to model user behaviors flexibly.

\subsection{Future work}\label{online}
Through these experiments, we fully understand the advantages and limitations of FE-TCM, and these limitations can further inspire some future work. For example, 1) For input features, we plan to expand click models by combining rich contextual information (e.g., dwell time, action type, vertical information, etc.), and making full use of more user behavior information (e.g., visual bias, mouse movement, etc.). 2) Many existing studies have shown that incorporating the embedding learned from graph structure can significantly improve model performance\cite{chen2020context,lin2021graph,jiang2018rin}. Graph structure must be exploited in the future to further improving session and click modeling. 3) In the users’ intention prediction, existing studies incorporate the pre-training method to calculate the topic relevance and assist intention prediction\cite{zuo2022improving}. Because this type of data contains query and document information, integrating topic relevance into click models can also be used as one of the directions to accurately predict users' click behaviors.

\section{Conclusion}
In this paper, we propose a new Filter-Enhanced Transformer Click Model (FE-TCM) for web search. We use the filtering algorithm in the field of signal processing, and ablation experiments prove that the learnable filters can effectively alleviate the influence of noise in sequence data. In addition, we model an attraction predictor and a examination predictor respectively, and apply Transformer to extract the dependence of different features in users behavior sequence. We have conducted extensive experiments on two real-world session datasets, and proves the superiority of our proposed model in click prediction task compared with the existing advanced click models.

\section*{Acknowledgments}
This work was supported by grants from Natural Science Foundation Project of Ningxia Province, China titled "User-oriented Multi-criteria Relevance Ranking Algorithm and Its Application (2021AAC03205)", Key R\&D Program for Talent Introduction of Ningxia Province China titled “Research on Key Technologies of Scientific Data Retrieval in the Context of Open Science (2022YCZX0009, 61862001)”, Starting Project of Scientific Research in the North Minzu University titled “Research of Information Retrieval Model Based on the Decision Process (2020KYQD37)” and North Minzu University Postgraduate Innovation Project (YCX22178, YCX22193).

\bibliographystyle{unsrt}  
\bibliography{references}

\end{document}